# Blastocoel morphogenesis: a biophysics perspective

Mathieu Le Verge-Serandour[1], Hervé Turlier[1]*

[1] Center for Interdisciplinary Research in Biology (CIRB), College de France, CNRS, INSERM, Université PSL, Paris, France.

* herve.turlier@college-de-france.fr

## Table of content

# Abstract

The blastocoel is a fluid-filled cavity characteristic of animal embryos at the blastula stage. Its emergence is commonly described as the result of cleavage patterning, but this historical view conceals a large diversity of mechanisms and overlooks many unsolved questions from a biophysics perspective. In this review, we describe generic mechanisms for blastocoel morphogenesis, rooted in biological literature and simple physical principles. We propose novel directions of study and emphasize the importance to study blastocoel morphogenesis as an evolutionary and physical continuum.

# Introduction

The blastula is the first identifiable stage of early embryo development that is common to many animals, and consists generally of a hollow epithelial layer enclosing a fluid-filled cavity: the **blastocoel**. Since the seminal work of Wolpert and Gustafson [1], blastocoel emergence is classically described in textbooks as the outcome of cell cleavage patterning. Yet this canonical view is challenged by the discovery of a diversity of mechanisms among animal species. Interestingly, these mechanisms are not necessarily exclusive and have remained rarely analyzed thoroughly from a physics viewpoint. In this review, we try to draw a generic overview of blastocoel formation from the available biology literature and we question the underlying mechanisms with a - sometimes original - physics perspective.

The blastula stage displays various configurations, not all of them including a fluid-filled cavity (see Box 1). Preceding gastrulation, blastulation prepares the embryo for the subsequent formation of primordial germ layers. For clarity, we introduce in Fig. 1 four generic classes of blastulae among animal embryos. Among them, the coeloblastula is prototypical of blastocoel formation, and will be the focus of this review; the discoblastula and periblastula have a large yolk region, that forms a physical barrier that generally prevents holoblastic (e.g. full) cleavages and leaves no or little space for a fluid-filled cavity, while stereoblastula remains densely packed morula-like balls of cells.

The blastocoel is the first cavity - or lumen - that appears in development, but it has remained little studied, except in early studies of sea-urchin and amphibian embryos [1], [2], [3]. In contrast to its apical counterpart [4], [5], [6], the blastocoelic lumen is located at the basolateral side of polarized cells (Fig. 2). This difference in localization is of importance, as the basal side of epithelial cells is generally enriched in cadherin adhesion molecules that favor close cell-cell contacts, while apical domains have repulsive electrostatic properties [7], favoring cell membranes repulsion and facilitating cavity opening [8]. One may hence expect distinct mechanisms for the formation of basolateral and apical lumens. *De novo* mechanisms for apical lumen opening and maintenance [4], [5] include tissue folding [9], fluid accumulation through an ion-generated osmotic gradient (zebrafish gut [10]), cavitation by cell apoptosis (breast tissue [11]), and vesicular trafficking (MDCK cysts [12] or Drosophila tracheal cells [13]). Interestingly, MDCK cells have the ability, to form either apical or basolateral cysts [14], depending on the environment where they grow (extracellular matrix or cellular medium), showing that apical and basolateral lumens can also share common morphogenetic processes, such as vectorial transport. For blastocoels (and more generally basolateral lumens), the reported mechanisms so far include surface plane divisions in sea urchins [1], [15] and many other marine invertebrates such as cnidarians (jellyfish [16], sea anemones [17]), water uptake triggered by an osmotic gradient in *Xenopus Laevis* [18], [19] and mouse embryos [20], [21], [22], [23], the fracturing of cell-cell contacts in cell aggregates in vitro [24] and several mammalian embryos [25], [26], [27], the hydraulic coarsening of micrometric cavities in mouse [25], and cell proliferation and spreading on the eggshell (e.g. zona pellucida) in marsupials [28]. In the next, we will illustrate each of these mechanisms in the model organisms where they were originally reported, and we will show that, rather than being exclusive, several mechanisms are generally at play and complement each other to ensure the proper morphogenesis of a blastocoel.

| BOX 1: **Different types of blastulae** |
|---|

The **blastula** (from βλαστος, the germ) follows early cleavages and precedes gastrulation [29]. Blastula can be classified into four general types, with a large diversity amongst each type:

The **coeloblastula** (from κοιλος, hollow) consists in a hollow sphere of cells enclosing the blastocoel cavity [30]. Coeloblastulae are found in amphibians (*Xenopus Laevis* [31]), cephalochordates (*Amphoxius* [32]), cnidarians (*Clytia hemisphaerica* [16]), echinoderms (*Sea urchins* [33], [34]), or therian mammals such as marsupials (*Monodelphis domestica* [35]) and eutherians (*Mouse* [25], *rhesus monkey* [26], *pig* [27]), where it is referred as a **blastocyst**.

The **stereoblastula** (from στερεος, solid or hard) consists of a solid ball of blastomeres [30] such as some sponges (*Petrosia ficiformis* [36]), nematodes (*C. Elegans* [37]), ascidians (*Phallusia Mammilata* [38], [39]), and is often found among spiral cleavages embryos such as mollusks (*Crepidula fornicata* [40]). Sometimes a very small or transient blastocoelic space emerges in the center, such as in some nematodes [37].

The **discoblastula** corresponds to a cap or a disc of blastomeres at the animal pole, above a yolk mass/cell, often produced by meroblastic cleavages in which the cells divide on the top of the yolk [30]. This type of blastula is typically found in avians [41], reptiles [42], fishes [43] and monotremes [44], [45]. There is no real blastocoel, but a slim subgerminal cavity between the blastoderm and yolk is classically observed in avians [41], [46].

The **periblastula** refers to an outer monolayer of cells enclosing an inner yolky mass, which may be seen as a substitute for the blastocoel cavity [30]. Such blastula may be found in centrolecithal eggs with dense yolk mass [47]. Cells are located at the periphery and form by superficial (meroblastic) cleavages, such that as *Drosophila Melanogaster* [48], [49] or holoblastic cleavages as in *Acari longisetosus* [50].

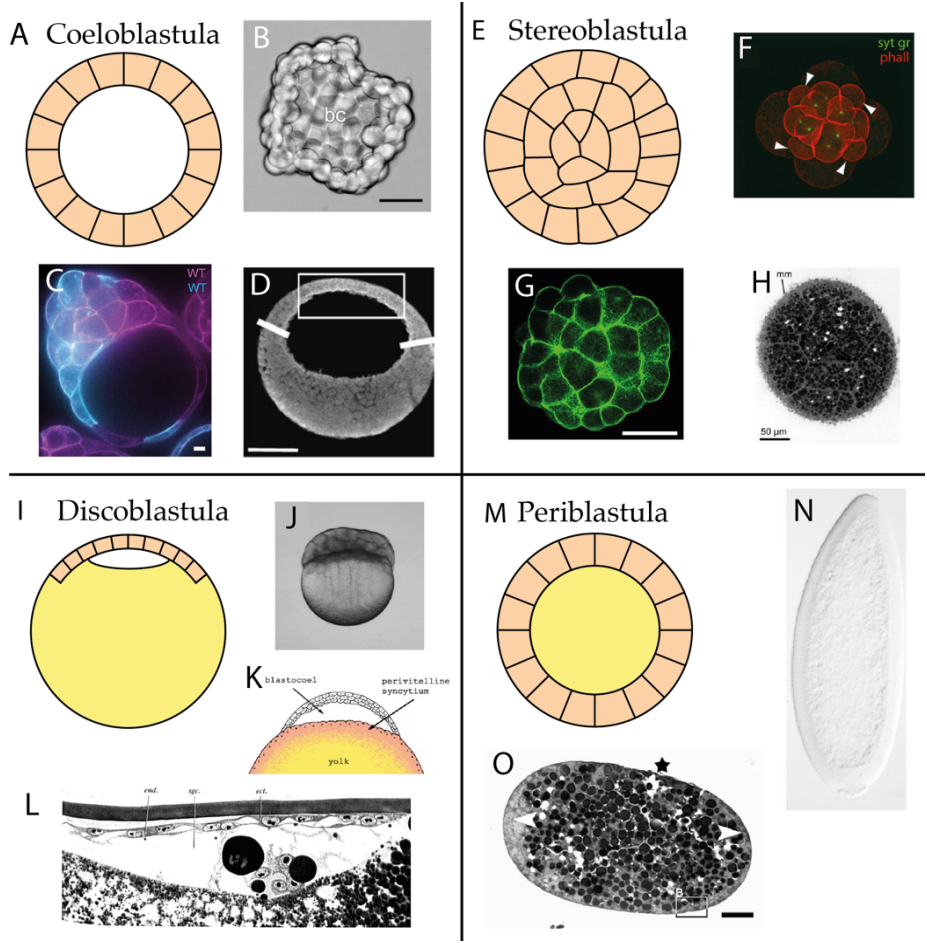

Figure 1. The four types of blastula. Blastomeres are represented in orange, yolk in yellow. In clockwise order: (A) *Coeloblastula:* (B) Cnidarian *Clytia Hemisphaerica* [16], (C) Mammals mouse [25], (D) Amphibian *Xenopus Laevis* [51]; (E) *Stereoblastula:* (F) Spiralia *Maritigrella crozieri* [52], (G) Hydrozoa *Gonothyraea loveni* [53], (H) Porifera (sponge) *Petrosia ficiformis* [36]; (I) *Discoblastula:* (J) Fish *Misgurnus anguilicaudatus* [43], (K) Avians [54], (L) Monotremes *Palyptus* [45]; (M) *Periblastula:* (N) Insect *Drosophila melanogaster* [48], (O) *Acaria longisetosus* [50].

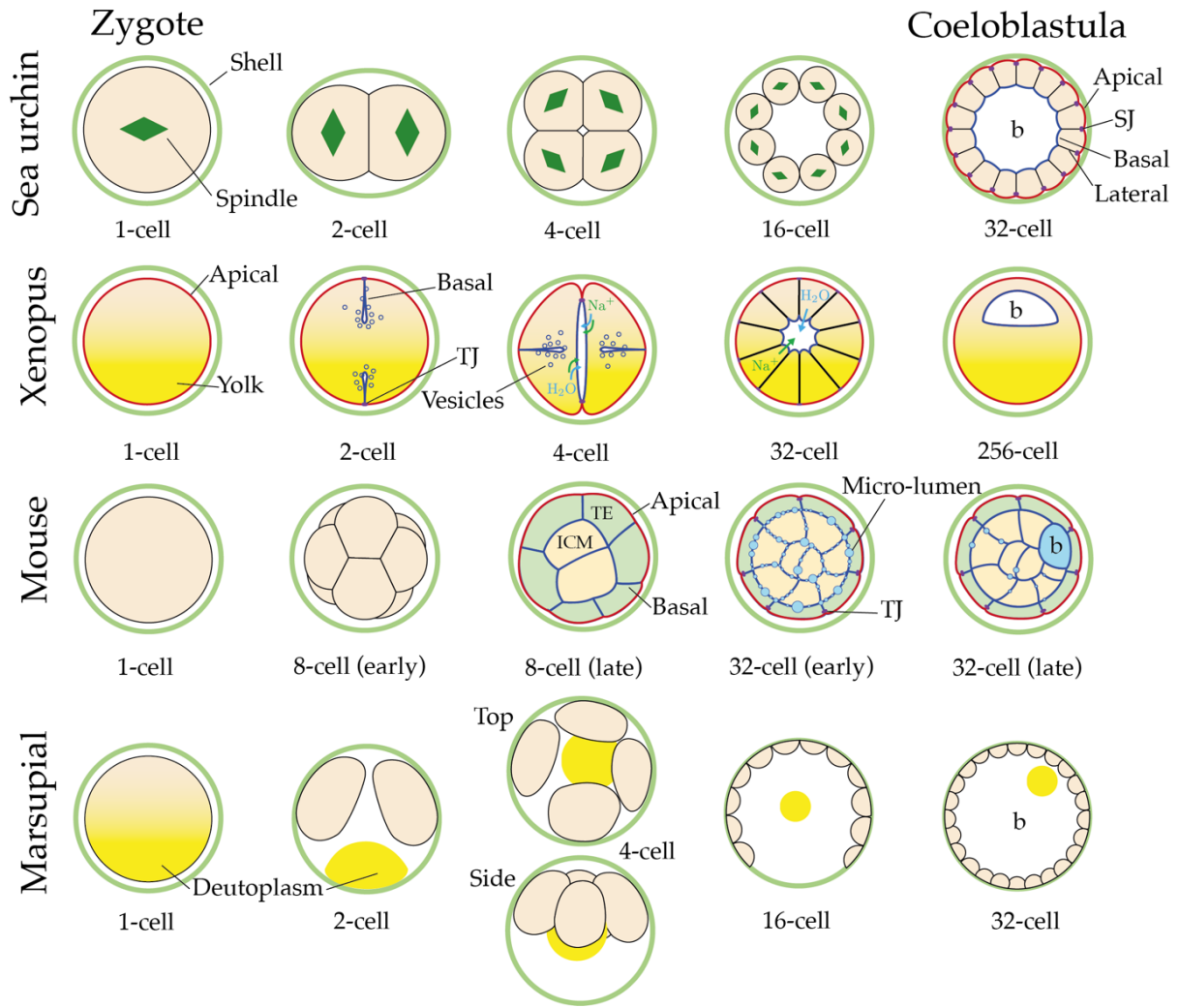

Figure 2. Four different reported modes of blastocoel morphogenesis. Sea urchin: surface plane divisions lead to a hollow cell layer surrounding a cavity [55]. Xenopus: the secretion of $Na^+$ through excretory vesicles and active pumping creates an osmotic gradient driving water uptake [18]. Mouse: the morula is fractured into myriad of micrometric lumens which coarsen into a single blastocoel cavity [25]. Marsupial: cell spreading and proliferation onto the shell (zona pellucida) leads to a single monolayer surrounding a single cavity, adapted from [56]. Apical (red), basal (blue), and lateral contacts are indicated; b: blastocoel, TJ: Tight Junctions, SJ: Septate junctions, TE: Trophectoderm, ICM: Inner cell mass.

# Mechanisms for blastocoel morphogenesis

## Sea urchin: surface plane divisions

Sea-urchin embryos, which belong to echinoderms together with starfish or holothurians, have been studied since the early 20[th] century [57]. This may explain why the predominant view for blastocoel formation is historically associated with this species [34]. Following fertilization, the embryo divides by equal cleavages along the embryo surface until the fourth cleavage, which is asymmetric and defines the animal-vegetal axis of the embryo [34]. The embryo is enclosed within a hyaline layer (Fig.2), which contributes to the close apposition of blastomeres at the 2-cell stage, the shape of the latter remaining otherwise more spherical [58]. This indicates that blastomeres do not compact naturally much to each other (see Fig. 3), leaving a small intercellular central space, that is amplified over successive divisions, to lead to a blastocoel cavity, that starts being distinguishable from the 16-cell stage onward [1]. The initial cavity expands and deflates rhythmically with cleavages, with a net volume increase over time [1], so that the embryo becomes ultimately a hollow sphere of epithelialized cells enclosing the blastocoel, in other words, a coeloblastula. The blastula is said to be formed between the 7th and 10th cleavages, as the permeability barrier is established around the 8th cleavage by septate junctions [59]. Two opposing models for the blastocoel formation have been historically proposed; the first model, supported by K. Dan, suggested that blastomeres adhesion to the hyaline layer and fluid accumulation, due to an osmotic gradient generated by secretion within the intercellular space, is at the origin of the blastocoel [33]. Interestingly, the 2-cell stage in some sea-urchin species (*Mespilia globulus* for instance) displays a large lens-shaped intercellular space suggesting an osmotically-generated hydrostatic pressure, that is inhibited by a hypertonic medium [60]. The second model, supported by Wolpert & Gustafson [1], proposed that adhesion between blastomeres and cleavages along the surface plane of the embryo (divisions called originally "radial" by the authors, who referred to division planes) were mainly responsible for the blastocoel formation [55]. Ultimately, the Wolpert & Gustafson's model was preferred, leaving however unexplained some early observations by K. Dan and others. Numerical simulations [15], [61], have since illustrated the potential of this canonical cleavage model to form a blastocoel, but with largely prescribed rules for divisions, adhesion, etc…, leaving many biophysical questions unsolved: What determines the orientation of divisions along the surface plane (cell geometry, adhesion, polarity cues)? What role is played by putative osmotic gradients disclosed in early and late blastocoel formation? Is the hyaline layer essential to blastocoel formation? According to Wolpert and Gustafson themselves: "It is by no means obvious why successive divisions should give rise to such a hollow sphere rather than a compact random clustering of cells." [1].

## *Xenopus*: osmotic fluid pumping

*Xenopus Laevis*, which belongs to amphibians, is another historical model system for the study of blastulation. The *Xenopus* egg reaches millimeter size and forms a multi-layered coeloblastula [62]. The outer apical surface is inherited from the egg, thus creating an early polarization of the embryo [63], [64]. A precursor for the blastocoel cavity is visible from the 2-cell stage onward [2], [3], and was described to be the result of basolateral membrane insertion by exocytosis of vesicles transported to the cytokinetic ring [65]. This is accompanied by the formation of tight junctions and adherens junctions as early as the 2-cell stage [66], [67], that become fully mature and seal the embryo around the 32-cell stage [66], [68]. In this embryo, it was early demonstrated that blastocoel expansion relies on osmotic forces [18], [66], which drive a net influx of water in the cavity [69]. Indeed, hypertonic media or the inhibition of the Na/K pumps by injection of ouabain in the nascent cavity prevents the formation of the blastocoel, the volume of which is metabolically controlled [18]. However, osmotic forces alone may not be sufficient to explain blastocoel emergence, as single blastomeres separated from the embryo at the 64-cell stage develop into an aggregate of cells that exhibit apicobasal polarity and can integrate into an 8-cell stage embryo, but do not form a blastocoel *de novo* [67]. This suggests that the establishment of a cavity may require initiation at early embryonic stages, starting from the 2-cell stage. Unlike in sea-urchin, where the blastula forms a hollow sphere, the blastocoel of the *Xenopus* is positioned toward the animal pole and breaks the embryo symmetry. Remarkably, the *Xenopus* blastula presents small interstitial gaps at multicellular junctions, where local adhesion is disrupted due to secretion of extracellular matrix, which is also found present in the blastocoel [70]. Interestingly, the absence of a large blastocoel by injection of ouabain does not impair embryo development into a tadpole in *Xenopus* embryos, which raises the important question of the role of the blastocoel in this species [18]. Blastocoel formation in *Xenopus* may hence result from a combination of several mechanisms, including cell division, hydro-osmotic pumping, and production of extracellular matrix. The relative roles and interplay of these mechanisms, as well as the initial nucleation process, remain however to be better described from a physics perspective.

## Mouse embryo: hydraulic fracturing and coarsening

Another well-studied blastula is the early mouse embryo. Mouse preimplantation development follows several morphogenetic steps [25], [71], [72], that lead to the formation of a so-called blastocyst. This structure, composed of an outer layer of epithelial cells, the trophectoderm, enclosing both the blastocoel and an inner cell mass lying on opposite sides [73], [74], needs to implant in the maternal endometrium to further develop. Before fertilization and until implantation, the embryo is surrounded by the zona pellucida (ZP), an outer shell of glycoproteins that is common to all mammals. Following fertilization, the mouse zygote cleaves and forms a tightly compacted morula at the late 8-cell stage (Fig. 2), thanks to an increase of cortical tension at the cell-medium interface of blastomeres [72] and the maintenance of a low tension at adhesive cell-cell contacts [75]. Apico-basal polarity emergence is concomitant [76] and triggers the appearance of tight junctions that become fully mature and effectively seal the embryo from the 32-cell stage onward [77], exactly when a blastocoel cavity appears and starts growing. Functional tight junctions are indeed crucial to blastocoel formation [78], as well as ion pumping through Na/K pumps located at the basolateral side of blastomeres [79] and ion transport by the Na+/H+ cotransporter (NHE-3) at the apical side of the embryo [80]. Vectorial ion transport and osmotic forces are therefore not specific to amphibian embryos, but how such processes lead to blastocoel emergence in tightly compacted morulae had remained unclear for years. New physical mechanisms were recently disclosed: a hydraulic flux generated osmotically by ion pumping in the intercellular space fractures cell-cell contacts into myriad of micro-sized cavities, or microlumens, that had yet already been observed by electron microscopy in monkey [26] or pig embryos [27]. This transient lumen network then coarsens through a process akin to Ostwald ripening, whereby smaller or more tensed cavities are more pressurized and empty themselves into larger ones by hydraulic exchange, ultimately leading to a single cavity [25], [81]. Blastocoel formation in mice, but also presumably in pig, bovine rabbit and human embryos, depends hence on functional ion transport through a polarized epithelium but is also characterized by hydraulic fracturing and coarsening, revealing the essential role of hydraulic and osmotic phenomena in early development.

## Marsupials: shell adhesion and cell proliferation

Marsupials are the closest relatives to eutherian mammals such as mice, but they retained some features of reptilian development, such as the conceptus coats and the presence of yolk (also called deutoplasm in this context) [82]. The composition of the multi-layered eggshell coat differs between species, but its generic structure is shared: an inner Zona Pellucida (ZP) is enclosed into a mucoid coat and surrounded by an outer shell coat [82]. Marsupials exhibit a specific mechanism for blastula formation: no morula nor compaction is observed [83], [84], but the blastocyst is formed through the proliferation of blastomeres onto the inner surface of the ZP [35], [85]. From the first to the fourth cleavages, the blastomeres exhibit low or no adhesion between themselves but adhere to the ZP [28], [83], [84], [86]. Adhesion to the ZP precedes cell-cell adhesion and is ensured either by club processes for *Trichosurus vulpecula* [87] or by a rim of zona for *A. stuartii* [84]. The yolk is extruded in the future blastocoel cavity over successive cleavages, thereby decreasing blastomeres volume [86]. At the four-cell stage, blastomeres lie well separated onto the ZP forming a cross-like shape [28].

Then, by cell proliferation and elongation, they progressively cover the ZP to form a unilaminar epithelial layer, enclosing the fluid-filled cavity [86]. Blastocysts can have different numbers of cells, around 32-cell for *S. crassicaudata*, *S. Macroura* [83], *A. Stuartii* [86], *T. vulpecula* [87]; 75-cell (bandicoots *Isoodon macrourus* and *Perameles nasuta* [88]); 100-cell (*D. viverrinus*) [89]. As soon as the blastocyst is formed, the embryo undergoes a large expansion through water uptake [90], suggesting vectorial ion transport after the late establishment of the epithelium. After completion, cells divide towards the center of the embryo by delamination to give rise to a new cell lineage [84], [90], [91].

## A physics perspective on blastocoel morphogenesis

After the biological review above, we try to provide generic physics perspectives on the reported mechanisms, starting with an overview of the main physical processes involved.

### Overview of the main physical processes involved

**Adhesion and cortical tension**

The shape of blastomeres is mainly determined by the mechanical forces exerted at the cell's surface, specifically adhesion and contractility [92] (Fig. 3). Contractility is generated by the actomyosin cortex and is controlled by the activity of myosin motors, which leads to an effective surface tension, often called cortical tension $\gamma_{cm}$ at free cell-medium surfaces. Adhesion contributes negatively to the effective interfacial tension $\gamma_{cs}$, and can be mostly attributed to cadherin molecules at cell-cell interfaces. In practice, the direct negative contribution stemming from cadherin bond energy, that we denote $\varepsilon_{adh}$, is often negligible compared to cortical tension. The lowering of tension at contacts is mostly mediated by signaling, which depletes the density of actomyosin material at adhering cortices by a factor $\Delta < 0$ (Fig. 3A) [93]. For a cell lying on a surface, its spreading may be described analogously to the one of a droplet through the balance of contact and cortical tensions only, $\gamma_{cs}$ and $\gamma_{cm}$ respectively, related to the contact angle $\theta_{cs}$ by the Young-Dupré equation (Fig. 3A). Cortical tension $\gamma_{cm}$ is generally considered uniform on the free surface of the cell in static situations. As $\gamma_{cm}$ increases, the contact angle $\theta_{cs}$ also increases. The minimal system to study the compaction of an embryo within a shell consists in two cells lying on a substrate (Fig. 3B). Two force balances are written, for the cell-cell contact and for the cell-substrate contact, which are described by two contact angles, $\theta_{cm}$ and $\theta_{cs}$ respectively, or equivalently by two compaction parameters, $\alpha_{cs} = \cos\theta_{cs}$ and $\alpha_{cm} = \cos\theta_{cm}$. These parameters tend to 0 if cells are compacted, and to 1 otherwise. Substrate spreading ($\alpha_{cs}$) and cell-cell compaction may compete with each other to control the cell's shape. Strong cell-cell compaction may impede blastocoel formation, as it possibly happens in most nematodes, that form a stereoblastula with an ephemeral and narrow blastocoel [94]. Substrate spreading and weak cell-cell adhesion on the contrary separate blastomeres that preferentially stick to the egg-shell, such as in marsupials, ensuring that no blastomere gets internalized [28], [44].

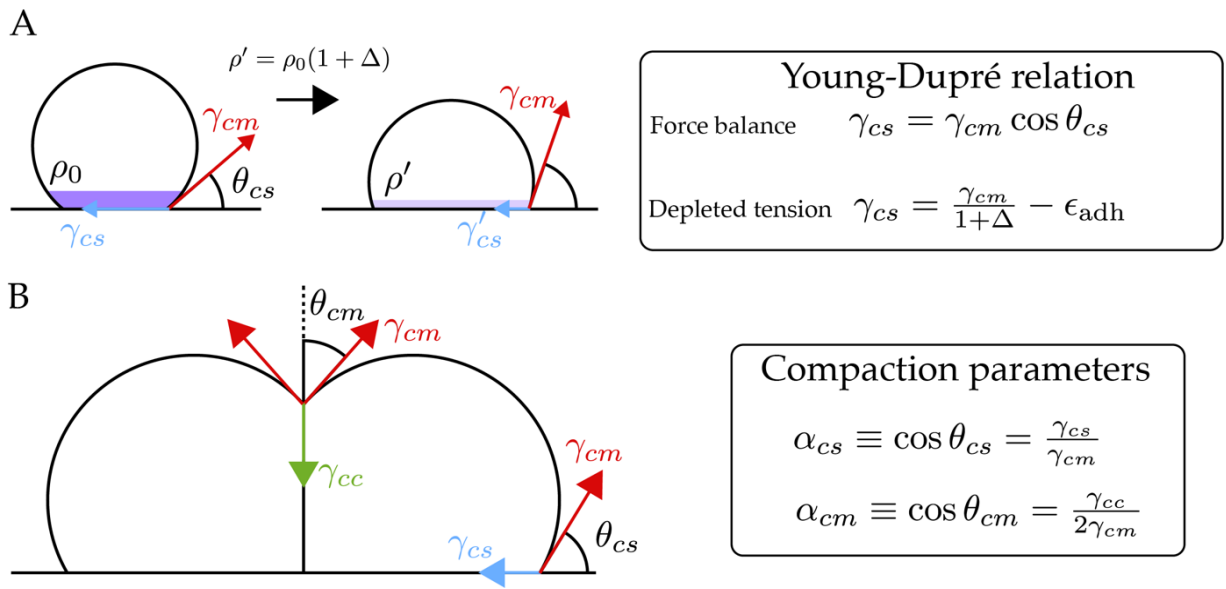

**Figure 3. Spreading and compaction of blastomeres. (A)** Spreading of a cell lying on a substrate, $\rho$ for cortical tension density, $\Delta$ for depletion factor. **(B)** Adhesion of two cells lying on a substrate with respective force balances.

**Control of cleavage patterning**

One of the first models of blastocoel formation was proposed by Wolpert and Gustafson in sea urchin [1]. The described mechanism relies primarily on cleavage patterning, whereby successive divisions have to be oriented tangentially to the embryo surface. Yet, what biophysical processes control and maintain the orientation of divisions along the surface plane in sea urchin and other species remains unclear. The cell cleavage plane is generically dictated by the positioning and orientation of the mitotic spindle. In early embryos, several cues may influence the location and orientation of the spindle [95]. The geometry of cells favors spindle orientation along the longest axis: this is the so-called Hertwig's rule [96]. But polarity cues, either cytoplasmic, such as yolk, or membranous/cortical such as apicobasal polarity [97], can also influence or direct the location and orientation of the spindle. Here we will not discuss the role of cell surface polarity cues, as they are often specific to each species. We rather discuss in Fig. 4A three generic physical processes: **i) adhesion to a substrate**, like an egg-shell, that can promote cell elongation along the embryo surface and planar divisions; **ii) cell arrangement**, which depends on both on cell-cell compaction and available space, controlling the aspect ratio of blastomeres; **iii) the role of yolk**, that may limit available space and favor a specific spindle location and orientation when present as a gradient in the cytoplasm.

**i) Shell adhesion** - Blastomere adhesion to an outer shell, like the hyaline layer in sea urchin [1] or the ZP in marsupials [87], can change their aspect ratio: specifically, the orientation of the longest axis of a single cell depends on the spreading of a cell onto the substrate (Fig. 4B). The degree of spreading or compaction, $1 - \alpha$, dictates not only the contact angle $\theta$ between the cell and substrate (see Fig. 3A) but also the overall cell aspect ratio $r = L/H$, where $L$ is the length, $H$ the height of the cell ($r = 1$ for a sphere, $r > 1$ for a flattened cell and $r < 1$ for a columnar cell). As soon as a single cell sticks to a substrate, its aspect ratio is larger than 1 (Fig. 4B). Thus, even the slightest adhesion is expected to orient divisions planarly according to Hertwig's rule, as this happens presumably in marsupial embryos. In sea urchin, cell-cell compaction, which promotes on the contrary columnar cell geometries (Fig. 4C), will enter in competition with a reported cell's spreading onto the hyaline layer.

**ii) Cell arrangement** - As cells divide within an egg-shell (with no adhesion to it), their shape is not only controlled by compaction (Fig. 3) but also constrained by the available space (Fig. 4C). For several cells confined within a shell, the longest axis depends on several parameters. We consider for simplicity blastomeres with flat lateral cell-cell contacts and with free apical and basal surfaces characterized by two curvatures, which allows us to define an aspect ratio $r = L/H$ (Fig. 4D). The expected division orientation is tangential (resp. transverse) to the embryo surface if $r > 1$ (resp. $r < 1$). We first consider in Fig. 4C an increase in cell number at constant cell height (left to right): this increases cell contacts, thereby reducing the embryo and cavity sizes, and drives a switch of division orientation from planar to radial, as cells become more columnar. If we consider an increase of cells contacts (or height) at constant cell number ($N_c = 10$ on Fig. 4C from bottom to top), this is also expected to reduce the cavity size and to drive a transition in division orientation from planar to radial. Thus, non-compacted embryos and little confinement are expected to promote cleavages along the embryo surface plane, while cells in compacted and confined embryos may preferentially cleave radially. Compaction and embryo confinement [98] are therefore key factors to quantify thoroughly to understand blastocoel formation in several species, including sea urchin.

**iii) The role of yolk** - Many fertilized eggs contain not only maternally deposited information required by the embryo to develop, but also nutrients in the form of yolk [30]. Yolk amount and distribution is another key factor to understand how cleavages will be oriented and completed [95]. In Xenopus, cleavages are unequal due to yolk granules that create a denser vegetal pole, that compete for space with the mitotic spindle. Cells at the vegetal pole become larger than those at the animal pole, where the blastocoel is located. The physical mechanism for the exclusion between the yolk and mitotic spindle remains however largely unexplored. In marsupials, yolk (or deutoplasm) may play another role, as its secretion reduces the volume of the blastomeres, contributing to the detachment of cells from each other and to their preferential adhesion to the shell [44].

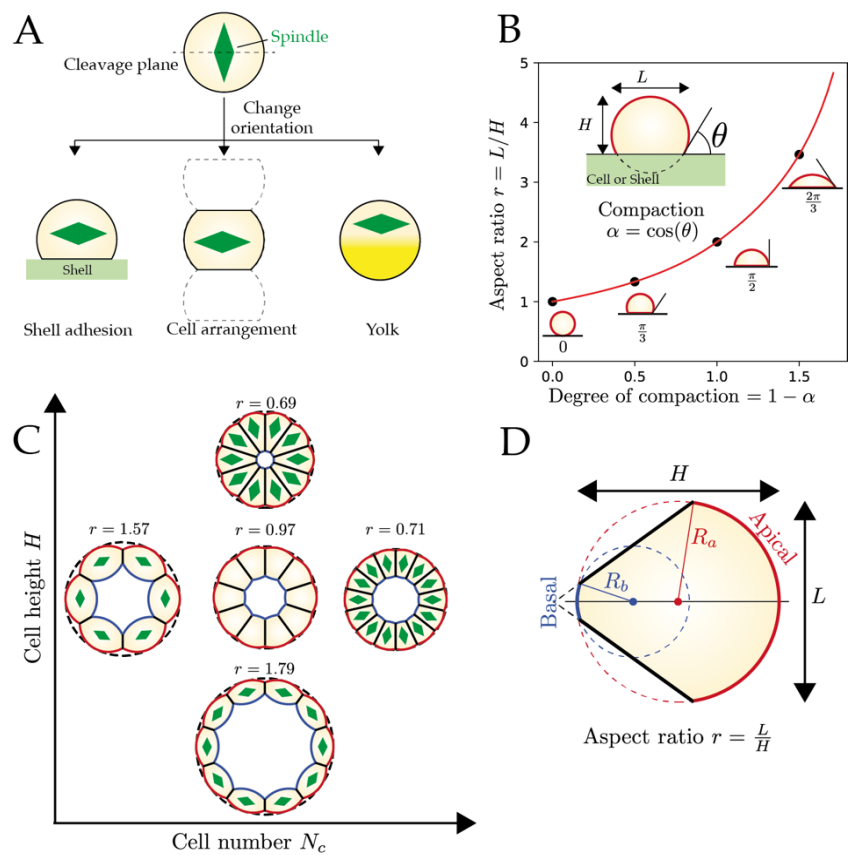

Figure 4. Control of cleavage patterning. (A) Mechanical cues for spindle orientation; the spindle is represented by a green diamond shape, and its orientation sets the cell division

axis. (B) Cell aspect ratio vs degree of compaction for a cell adhering to a shell, at constant cell volume; four states of compaction are depicted, the contact angle is given below each. (C) Change in spindle orientation (green diamond) expected from the long-axis (or Hertwig's) rule, when the cell number or height is changed at constant total cell volume. (D) Definition of our blastomere geometrical measures.

**Lumen interaction: hydraulic coarsening and coalescence**

The formation of the blastocoel can involve the prior emergence of multiple smaller cavities or lumen. But how such smaller cavities interact and evolve to form a single final lumen has remained little studied. In physics, coarsening phenomena, such as Ostwald ripening [99], describe the collective dynamics of a population of compartments, originating generally from the coexistence of two immiscible phases, and that are pressurized by a surface tension. One signature of coarsening is that the total number of compartments decreases but their average size increases through mass exchange, until a single droplet-like entity remains. This mass exchange is generally driven by surface tension, which creates Laplace's pressure gradients among compartments, but may be mediated by distinct mechanisms: diffusion, coalescence or more rarely hydraulic flows. Recently, we revealed a coarsening process in mouse embryos, that relies on hydraulic exchange to form the blastocoel [25]: small lumens created at cell-cell junctions by osmotic pumping remain connected through intercellular bridges and are hence able to exchange luminal fluid (Fig. 5A). Cortical tension leads to a hydrostatic pressure $P_i$ in cavity $i$, given by Laplace's law as a function of the surrounding cell pressure $P_0$, the radius of curvature $R_i$, and the surface tension $\gamma_i$ of the cavity (Fig 5A). Any difference in cavity size or tension generates a pressure gradient between two cavities, thereby driving a hydraulic flux. The flux from cavity 2 to cavity 1 can be expressed as $J_{2\to1} = (P_2 - P_1)/\Re_{12}$, where $\Re_{12} = l_{12}/\kappa_v$ is a hydraulic resistance with $l_{12}$ is the length of the connecting bridge and $\kappa_v = e_0^3/12\eta$ a Poiseuille's coefficient that accounts for the fluid viscosity $\eta$ and the bridge transverse thickness $e_0$ (here in 2D). When multiple cavities are interconnected, the system has only one stable state, where a single cavity is left, containing all the fluid. In the mouse embryo, the surface tensions $\gamma_i$ depend on the fate of cells, thereby biasing fluxes towards regions of lowest tension (or mechanical resistance). Cells in the inner cell mass (ICM) are expected to have higher contractility than outer trophectoderm cells (TE). At TE-ICM contacts, cavities are indeed asymmetric, bulging towards the TE, indicative of lower surface contractility at the TE side. This tension asymmetry between TE and ICM directs hydraulic flows towards low contractility cell-cell contacts and breaks the radial symmetry of the blastocyst to position the blastocoel at the TE-ICM interface [25]. As time grows, the total number of micro-lumens decreases. As expected for coarsening phenomena, one can predict a dynamic scaling law for the number of cavities as function time $N(t)\sim t^{-2/5}$ [100] (here for a 1D chain of 2D cavities). For higher dimensions, such as the 3D cavities lying at 2D interfaces of the mouse embryo, one may generalize the result and predict a scaling law $N(t)\sim t^{-3/4}$ [100], [101], that remains to be verified experimentally. This dynamic scaling law, and the self-similar nature of the underlying cavity size distribution in time, are reminiscent of coarsening in dewetting films [102] [103]. Interestingly, one finds that osmotic gradients play a negligible role in the lumen coarsening process [100]. Similarly, the permeation of water and solutes across the cell membrane, which may screen pressure and concentration gradients between cavities [100], [104], does not affect the scaling exponent, although a strong screening of pressure gradients shall lead to an overall collapse of cavities. The estimate of the pressure screening length in the mouse embryo leads to a value close to the embryo size (80μm): this means that the continuity of pressure gradients is ensured across the embryo, preventing the development of several blastocoels. A sustained active pumping of osmolytes, for instance by Na/K-ATPase pumps [79] may, on the contrary, largely influence the dynamics by enhancing cavity growth and driving a switch to a coalescence-dominated regime, characterized by a different scaling law $N(t)\sim t^{-1}$ for a 1-dimensional chain. In the mouse embryo, coalescence events of micro-lumens have been reported [105], but coalescence does not seem to be the dominant mechanism of lumen interaction in this context [25]. However, in a similar way as for contractility, patterning pumping within the embryo may be enough to statistically bias the position of the blastocoel, which will preferentially form where smaller cavities grow faster [100]. Further studies will be needed to account for electro-osmotic potentials generated by the charges of ions, such as sodium, potassium and chloride, which can furthermore take distinct transport routes (Fig. 5A). In particular, chloride is thought to be exchanged through tight junctions in mouse blastocysts, potentially creating a paracellular current [106]. Preceding coarsening, the mechanism for micro-cavity/lumen formation has not been fully elucidated yet. In the mouse blastocyst, it is suggested that cell-cell contacts undergo hydraulic fracking [107], by injection of pressurized fluid into the intercellular space, following ionic gradient establishment [25]. Overall, these processes highlight the essential roles of hydraulic fluxes as fundamental forces to shape embryos during their development [108], [109].

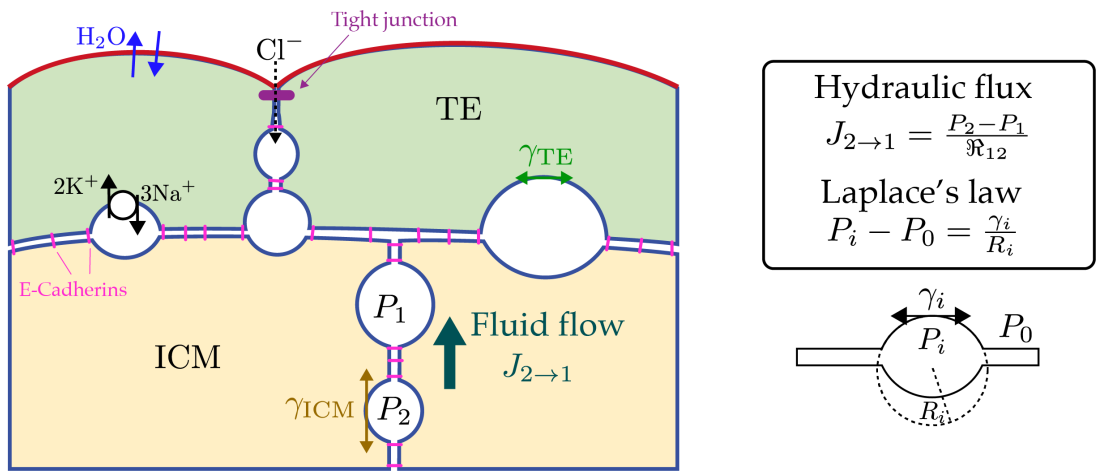

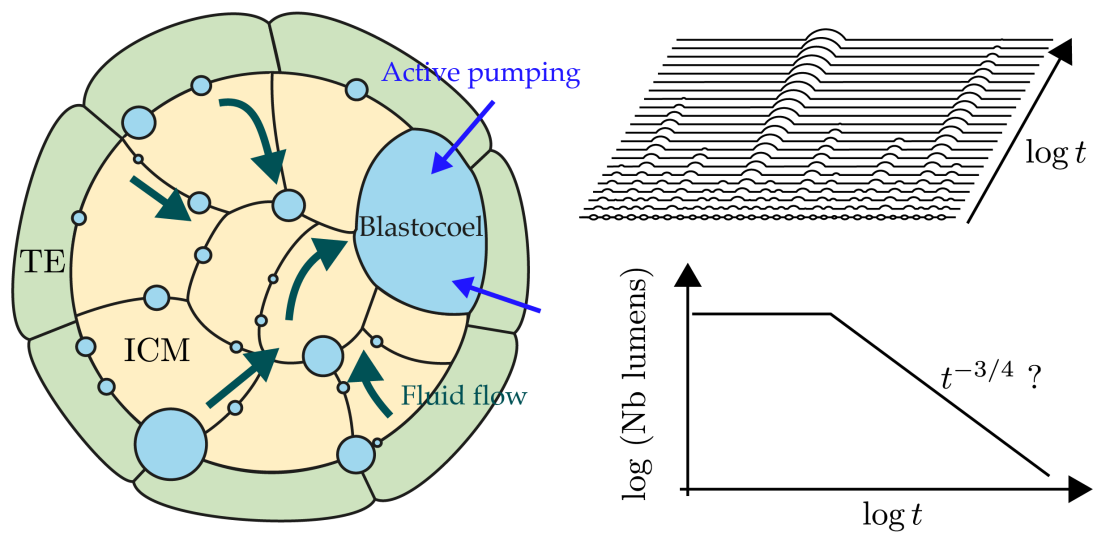

Figure 5. Hydraulic coarsening. (A). Cell-cell contact within a mouse embryo and geometry of a single lumen. TE: Trophectoderm, ICM: Inner Cell Mass. (B) Left: schematic of a mouse blastocyst with multiple micro-lumens and fluid flows. Right: chain of micro-lumens in time, and their number as a function of time.

### Polarized ion transport and volume regulation

Just as cells control their size, the coeloblastula may control the blastocoel size by passive osmotic transport, active ion pumping or mechanical forces [110]. To physically illustrate these effects, we simplify the description of a coeloblastula as a cavity surrounded by a single cellular compartment, that may be viewed as a continuous and sealed epithelial layer (Fig. 6A). Epithelial cells have specialized fluid and solute transporters with asymmetric distributions. The Na/K pumps, importing 2K$^+$ within the cell and secreting 3Na$^+$ into the intercellular space, are present at the basolateral membrane in mouse [79] and *Xenopus* [68], while Na/H transporters, importing Na$^+$ within the cell and secreting H$^+$ toward the external medium, are present at the apical membrane of the mouse embryo [80]. The selective transport of charged ions through semi-permeable membranes generates electro-osmotic potentials that cells use to control their volume [111], [112]. To simplify the physical picture, we consider here only one single uncharged solute. The blastocoel is characterized by its radius $R(t)$ and concentration $C_{in}(t)$ and the medium by its constant concentration $C_{out}$. We neglect concentration changes in the cell, which is modeled via a permeation $\lambda_v$ for water, and $\lambda_s$ for solutes, and a surface tension $\gamma_{epi}$. The dynamics of volume change in the cavity are proportional to water permeation and given by the competition between the hydraulic pressure difference, $\Delta P = P_{in} - P_{out}$, and the osmotic pressure difference $\Delta \Pi = \mathcal{R}T(C_{in}(t) - C_{out})$ (Van't Hoff relation, where $\mathcal{R}$ is the perfect gas constant and $T$ the temperature). Typical hydrostatic pressures generated by cortical tensions ($P \sim 10^{2-3} Pa$ [113]) are negligible compared to absolute values of osmotic pressures ($\mathcal{R}TC \sim 10^5 Pa$), but not to the relative difference. Indeed, when the two pressure differences equalize ($\Delta \Pi \sim \Delta P$), the volume reaches a steady state. It is important to note that, because of this large difference between orders of magnitude between absolute values of osmotic and hydrostatic pressures, the volume is generally controlled mainly by the osmotic imbalance: a change in Laplace's pressure, mediated for instance by a drop in cortical tension, may affect the final volume by less than 1%, which would be experimentally undetectable [111]. When the volume changes, so does the concentration within the cavity. If solutes can be exchanged, with a permeation $\lambda_s$, the evolution of solute concentration is controlled by the chemical potential difference $\mathcal{R}T log(C_{out}/C_{in})$ between the medium and cavity, and may increase via active pumping at a rate $j_s$. In the limit of fast solute relaxation, the active pumping $j_s$ equates to the outward osmotic flux $\lambda_s \mathcal{R} T log\left(\frac{C_{in}}{C_{out}}\right) \sim \lambda_s \mathcal{R}T(C_{in} - C_{out})/C_{out}$, such that the cavity will expand above a pumping threshold $j_s^* = \lambda_s(P_{in} - P_{out})/C_{out}$, and will shrink until disappearance below it (Fig. 6A). Extensions of such a simple single compartment model can account for larger stability regions of the volume as function of the pumping rate, first by including ion charges and the electro-osmotic membrane potential [111], [114]. Refinements may also account for cell division and oscillations that may generate cavity collapse [113], and short timescale cortical tension dynamics [104]. However, very few models have modeled the coupled volume control of cells and cavity as separate compartments [115]. While nothing is expected to prevent the collapse of the cavity, its expansion may be buffered mechanically by an external shell, whose resistance may exert a hydrostatic pressure of much higher amplitude than the one generated by cortical or epithelial tension (Fig. 6B). As an example, the zona pellucida (ZP) of mammalian embryos is often described as an elastic material [116]. One can define the tension of the shell $T_{shell}(R) = h(R)\sigma(R)$ as a function of its radius $R$, its thickness $h$ and the tangential stress $\sigma$ (Fig. 6B). At first approximation, the stress-strain relation for a simple elastic material may be taken as linear, $\sigma(R) = E(R - R_0)/R_0$, with $R_0$ is the resting radius of the shell and $E$ its Young's modulus, evaluated around 100 kPa after fertilization [117]. Note that if the shell is considered incompressible, its thickness will decrease as the radius increases, which may lead to an hyperelastic behavior at large deformations [116]. Shell tension adds up to the epithelial tension $\gamma_{epi}$ in mechanical balance, increasing Laplace's pressure difference well beyond hundreds of Pascals as the cavity grows [116]. A stable steady-state volume may hence, in this case, be attained if the pressure difference reaches the critical value $P_{in} - P_{out} = j^* C_{out}/\lambda_s$.

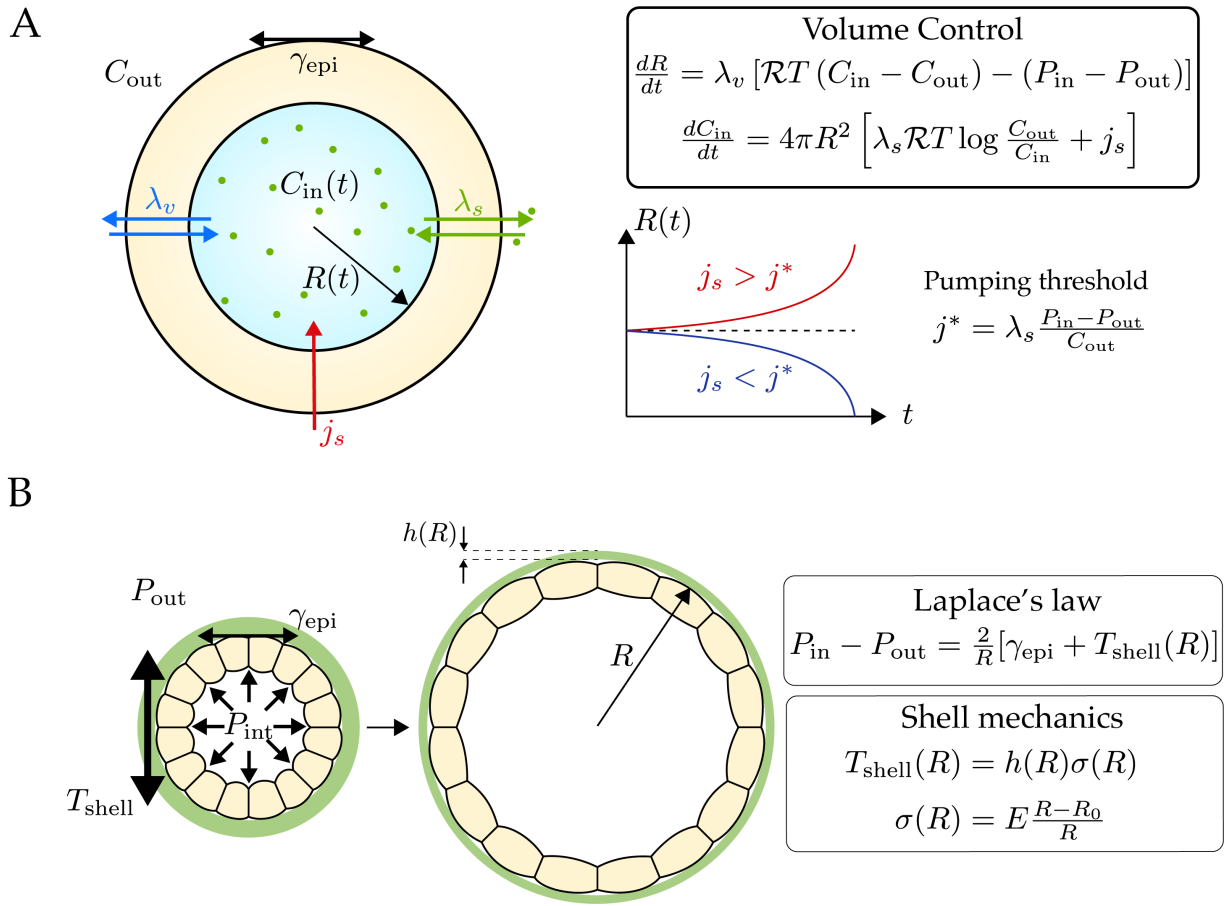

Figure 6. Polarized transport and osmotic volume control. (A) Model for the volume control of a cavity within a cell. (B) Mechanical contribution of the shell to the tension.

## Remaining gaps and novel perspectives

### Epithelialization and vectorial transport

The epithelium is one of the first of cell specialization during embryogenesis and separates the internal medium of an embryo from the exterior [118]. Epithelial cells are characteristic features of metazoa [119], [120]. They form layers of polarized cells, oriented along an apical-basal axis, and are kept together through adherent junctions and sealed by tight junctions in vertebrates [68], septate junctions in invertebrates [118], that prevent water leaking and selectively control the diffusion of solutes [121]. We refer to epithelium establishment by distinguishable apicobasal domains and by the presence of mature tight/septate junctions. Because of its sealing and transport properties, epithelium establishment is paramount to understand the mechanisms of blastocoel formation: depending on the timing of such establishment relative to other morphogenetic events such as compaction or shell adhesion, one particular mechanism may be preferred from the variety of mechanisms described above. While some recent work in mouse embryos described the extension of an apical domain as key to *de novo* polarization [122], we lack a clear biophysical picture of the mechanisms behind epithelialization, including the control of vectorial ion transport and tight junction maturation, which may be essential for blastocoel maintenance [123].

### Physical confinement and substrate adhesion

Unlike eutherian embryos, marsupials do not form a morula but adhere separately to the surrounding ZP, which is required for normal cleavages [35], blastocyst formation [86], and survival of the embryo [85]. Yet, a clear physical picture of blastocyst formation supported by shell adhesion is missing so far. More generically, the role of spatial confinement on blastocoel formation has been little investigated.

**Timing of cell division**

Timing of division may be as essential as the spatial patterning of cleavages for blastocoel morphogenesis. A striking example is ascidians embryos, which present invariant cleavages along the embryo surface [38]. In this species, cell cycle asynchrony emerges at the 16-cell stage, whereby the eight cells of the vegetal part divide first, such that the embryo reaches a 24-cell stage, followed by the eight cells of the animal part to reach the 32-cell stage [124]. This asynchrony remains until a 44-cell stage, since the two blastomeres of the germline divide apart [125]. Interestingly, while the wild-type embryo forms a compact stereoblastula, ablation of the contraction pole resynchronizes the cell cycle and leaves an intercellular space in the center of the embryo, that grows into a blastocoel in a seemingly similar fashion as in sea urchin [38], [125].

**A role for blastocoel emergence?**

An important question remains largely unanswered, in the literature and throughout this review: the role of the blastocoel. Blastocoel morphogenesis precedes gastrulation, and the formation of a hollow structure surrounded by a single cell monolayer surely prepares the embryo for subsequent folding and invagination of future primary germ layers. It also allows to clearly define inside and outside compartments in the embryo. From a mechanical perspective, it may be easier to deform a single monolayer than a bulk aggregate of cells. Yet, in many organisms, gastrulation operates fine without a blastocoel (or yolk cell). Most nematodes, for instance, are described as stereoblastula during their development, and while the most studied species, *C. Elegans*, does form a very small and transient cavity, it rapidly disappears and does not play a role in gastrulation [37]. Remarkably, some nematodes do form a distinct and prominent blastocoel, such as *Tobrilus stefanskii* at the 16-cell stage [94], [126], [127]. Another possible role for the blastocoel is to provide a simple mean for cell interactions via a morphogen, that should freely diffuse within the fluid-filled cavity [128]. On the contrary, it may provide a barrier between cells, to prevent direct biochemical interaction through cell-cell contacts. To better determine the intricate coupling between blastocoel morphogenesis and fate patterning, future studies will therefore need to combine realistic physical modeling with precision perturbation of blastocoel formation, possibly by physical means (osmotic chocks, mechanical compression…).

# Blastocoel morphogenesis: an evolutive continuum?

Metazoans are multicellular organisms, organized with several germ layers, composed of specialized cells [30]. The idea that early steps of metazoan emergence correspond to the early stages of embryonic development is an old idea [129], and even though it has been refuted multiple times, it has impacted the thoughts of early embryonic development [29], [130]. It has also been proposed that the emergence of metazoans is linked with the origin of epithelia [118], and those primeval metazoans might have exhibited three novelties: (i) assembling multicellular aggregates, (ii) establishment of occluding junctions, (iii) polarized transport [120]. Whether of the evolutionary origin or not, these similarities are worth exploring to decipher the order of events that may lead to the establishment of a fluid-filled cavity within the embryo, and how to conserve it during development.  The structure composed of a cellular compartment enveloping a cavity is reminiscent of colonies of unicellular protozoans, assembled to form a collection of cells surrounding an empty space [130]. Choanoflagellates for instance consist of a sperm-like cell, whose apical flagellate is surrounded by a "collar" of microvilli forming the collar complex [131]. Either by aggregation or cell division with the apicobasal plane, choanoflagellates may form multicellular structures [131]. Specifically, the model species *S. rosetta* forms rosette-like colonies - in the form of a hollow sphere of cells - whose integrity is ensured by basal extracellular matrix (ECM), intercellular bridges, and filopodia [132]. Moreover, rosette colonies significantly upregulated tight junctions and cadherin molecules, responsible for cavity sealing and cell-cell adhesion respectively [133]. Biophysical simulations have shown the importance of the ECM and its physical and mechanical properties to ensure the formation and stability of rosettes instead of other structures [134]. Acquisition of multicellularity through the aggregation of unicellular organisms is not the only hypothesis for metazoan formation. Another hypothesis suggests that multicellular systems formed through multi-nucleated syncytia that later acquired cytokinesis, a development closer to the one of arthropods forming a periblastula. Such example is found in the protozoa ichthyosporean *Creolimax fragrantissima*, which develops through multinucleated syncytium, with nuclei surrounding a central vacuole, a structure reminiscent of the coeloblastula. The vacuole is later filled with cytoplasm and the colony divides into single motile cells [135]. While these morphogenetic events are not enough to qualify protozoa colonies as metazoan, because they lack establishment of an axis of symmetry for the body plan and specialization of interdependent cells [30], this shows that the establishment of a small cavity in protozoans already had most of the physical ingredients present in the main mechanisms we discussed for blastocoel formation.

# Conclusion

Blastulation might not be "the most important time in our life", to refer to the famous quote of Lewis Wolpert on gastrulation [136], but it remains the very first morphogenetic event common to most animals. Surprisingly, it remains much less explored from a biophysical perspective. Maybe because the morphological changes undergone by embryos at this primitive stage of development are less striking? We argue that blastula formation, and by extension blastocoel morphogenesis in coeloblastula, reveals nevertheless in a very short amount of time, a wide diversity of seminal biophysical phenomena, many of them remaining to be revisited quantitatively, starting with historically-studied species. We tried, in this review, to provide a transverse picture of blastocoel formation across species, together with physical perspectives. We hope we could convince the reader that for many of them, they are not necessarily exclusive, but may, on the contrary, complement each other, or at least constitute redundant fail-safe mechanisms to ensure the proper formation and localization of the first fluid-filled biological cavity. We hope such a view may call, more generally, for a *physical continuum* approach of morphogenesis, whereby a single physical process may generally not be sufficient to provide a complete and fair picture of a biological phenomenon. Blastocoel emergence requires the consideration of as many prototypical physical concepts as surface tension, adhesion, hydraulic flows, osmotic volume control and mechanical constraints, concepts which operate together at later embryonic stages, presumably in an even more intricate manner. Integrating several physical processes in a single model is however often not an easy task, and will require further progress in 4D numerical simulations. While as physicists, we like to draw a simple physical picture of biological phenomena, blastocoel morphogenesis proves to us that the physics of development, even on one of the simplest multicellular system, is nothing but simple. From our review, it appeared that one key question remains broadly unanswered: what is the role of the blastocoel? While it may be argued that it provides the necessary space and mechanical compliance for subsequent tissue folding during gastrulation, these arguments remain hypothetical and many embryos do gastrulate without the need of a blastocoel. To tackle such questions, morphogenesis may largely benefit from evolutionary approaches. For blastula formation, this goes back to the emergence of multicellular assemblies in protozoans: seeing early embryo development as an *evolutionary continuum* could provide inspiring directions to address key phenomena such as polarization, adhesion and epithelialization. We hope these thoughts may lead to the emergence of a new field at the interface of physics and evolutionary morphogenesis.

# Acknowledgments

We thank S. Frankenberg, A. McDougall and R. Dumollard for fruitful discussions and comments on the manuscript. We thank F. Delbary and members of Turlier labs for their support. This work was supported by the CNRS through the ATIP-Avenir program, by Collège de France, by the Foundation Bettencourt-Schueller and by the EMBRC-France. This project has received funding from the European Research Council (ERC) under the European Union's Horizon 2020 research and innovation programme (Grant agreement No. 949267).

## Methods

The Jupyter notebooks and equations used to generate figures can be found here: https://github.com/VirtualEmbryo/blastocoel_review

## Lexicon

- **Adherens junction**: type of adhesive cell junction composed of transmembrane adhering molecules, linked to the actin cytoskeleton [34].
- **Animal pole**: in yolky cells or embryos, designate the hemisphere that is away from the yolk [34].
- **Apical**: side of a polarized epithelial cell that faces the exterior. Usually rich in microvilli and proteins specialized in exchanges with the exterior.
- **Apoptosis**: type of programmed cell death, in which the cell commits "suicide". This death does not damage other surrounding cells, unlike necrosis. [34]
- **Basolateral**: side of a polarized complementary to the apical. Composed of the lateral plus the basal faces of a cell.
- **Blastomere**: any of the cells formed by the cleavages of the fertilized egg [34].
- **Blastulation:** morphological step of the embryonic development that produces a coeloblastula, a layer of blastomeres surrounding a fluid-filled cavity. Also referred as cavitation.
- **Cadherin**: a family of cell-adhesion molecules [34].
- **Cavitation**: the process by which the blastocoel appears in the embryo.
- **Centrolecithal**: egg in which the yolk is located in the center [30].
- **Cleavage**: cell division without volume growth. Cleavages occur during early embryonic development [34].
- **Club process**: cell protrusion on the apical domain shaped as a club, contributing to adhesion [87].
- **Columnar cell**: column-shaped cell, with larger lateral side than basal and apical sides, typically in epithelia.
- **Compaction**: in the mouse embryo, the stage at which the blastomeres flatten by increasing their cell-cell contacts, and the overall embryo becomes rounder [34].
- **Contractility**: refers to the ability of a cell to change its volume by contraction.
- **Cortex**: of actomyosin. A dense network of actin polymers and myosin molecular motors beneath the cell membrane, that ensures the contractility of cells and is responsible for the cell surface tension.
- **Cyst**: an empty pocket within a tissue, that may contain air or fluids, and can generate tumors.
- **Cytokinetic ring**: the contractile ring of actomyosin that appears during cytokinesis and is responsible for splitting the mother cell into two daughter cells.
- **Deutoplasm**: secondary cytoplasm of eggs, distinct from the nutritive yolk. Its precise role is not yet elucidated [56].
- **Epithelium**: a sheet of polarized cells tightly bound together by adhesive cell junctions [34].

- **Eutherian**: clade of the Therian, distinct from Metatherians (Marsupials). They form a blastocyst with inner cell mass, the embryo implants early in development with a placenta. Eutherians encompass mouse, human, pig, etc. [56]
- **Exocytosis**: active transport of material outside the cell, by the fusion of vesicles with the plasma membrane. Opposed to endocytosis.
- **Extracellular matrix**: an extracellular network of molecules, that provide them mechanical and chemical support.
- **Gastrulation**: process in which tissues (endodermal and mesodermal) move from the outside to the inside of the embryo, and give rise to the internal organs [34].
- **Germ layer**: primary types of cellular tissues in the early embryo. The most common ones in animals are the ectoderm, mesoderm and endoderm [34].
- **Holoblastic**: cleavage with complete cytokinesis, in which the cleavage plane passes throughout the entire cell [30].
- **Hyaline**: in echinoderms, the external shell surrounding the embryo [58].
- **Inner Cell Mass**: a group of cells in the blastocyst of mammalian embryos that will give rise to the embryo proper. These cells are surrounded by the trophectoderm cells and facing the blastocoel after cavitation [34].
- **Invagination**: local deformation of tissue through the bending and folding of a layer of cells, such as in gastrulation. [34]
- **Lumen**: empty space in a tissue, often in a tubular structure such as blood vessels or intestine.
- **Meroblastic**: cleavage with uncomplete cytokinesis, in which the cleavage plane stops and the daughter cells are not fully separated [30].
- **Metazoa**: clade that designates the animals, or multicellular organisms.
- **Mitotic spindle**: assembly of microtubules in the shape of a spindle that pulls apart the chromosomes into daughter cells during mitosis [34].
- **Morula**: early stage in a mammalian embryo when cleavage has resulted in an aggregate of cells. The embryo looks like a blackberry [34].
- **Mucoid coat**: an acid-glycoprotein produced by the oviduct of marsupials, that provides nutrients and block polyspermy. It surrounds the zona pellucida and is enveloped in an outer shell coat [85].
- **Na/K pump**: or Na/K-ATPase, a specialized enzyme that at the cost of one ATP molecule, exports three sodium ions outside the cell, and imports two potassium ions within. It is a marker of basolateral cell domains.
- **Ouabain**: a cardiotonic steroid and inhibitor of the Na/K pump.
- **Protozoa**: the term that designates unicellular eukaryotes.
- **Semi-permeable**: the ability of a membrane to allow the passage of a single chemical species, and to block others. It typically let the solvent pass and retains solutes.
- **Septate junction**: intercellular junctions of non-chordate invertebrates, that act as a permeability seal. Their equivalent is the tight junctions in vertebrates [59].
- **Subgerminal**: (space) The cavity enclosed between the yolk and the blastoderm in monotremes and avians [34].
- **Syncitium**: multinucleated cell [34].
- **Tight junction**: a type of adhesive cell junction that binds epithelial cells tightly together to form an epithelium and seals off the environment on one side of the epithelium from the other [34].
- **Therian**: subclass of mammals, that encompass both Eutherians and Metatherians (marsupials). Therians do not lay shelled eggs, contrary to Prototherians (Monotremes)
- **Trophectoderm**: the outer layer of cells in the early mammalian embryo, consisting of a squamous epithelium that surrounds the inner cell mass and the blastocoel. Trophectoderm gives rise to extra-embryonic tissues and the placenta [73].

- **Vegetal pole**: in yolky eggs or cells, the hemisphere that is rich in yolk [34].